\documentclass[a4paper,12pt]{article}

\usepackage[english]{babel}
\usepackage{hyphenat}
\usepackage{bm}
\usepackage{amssymb}
\usepackage{amsmath}
\usepackage{amsfonts}
\usepackage{dsfont}
\usepackage{amsbsy}
\usepackage{graphicx}
\usepackage{authblk}
\usepackage[width=0.9\textwidth]{caption}
\usepackage[colorlinks,citecolor=blue,urlcolor=blue,linkcolor=blue]{hyperref}

\newcommand{\slsh}[1]{\not{\hbox{\kern-2pt${#1}$}}}

\newcommand{\ba}[1]{\begin{eqnarray} \label{#1}}
\newcommand{\ea}{\end{eqnarray}}

\def\bea{\begin{eqnarray}}
\def\eea{\end{eqnarray}}
\def\bqu{\begin{quote}}
\def\equ{\end{quote}}

\newcommand{\newc}{\newcommand}
\newc{\ra}{\rightarrow}
\newc{\lra}{\leftrightarrow}

\newc{\sm}{Standard Model}
\newc{\smd}{Standard Model}
\newc{\barr}{\begin{eqnarray}}
 \newc{\earr}{\end{eqnarray}}

\def\LSP {$\chi^0_1$}
\def\stau{$\tilde{\tau}_1$}
\def\staunu{$\tilde{\tau}_1 -\tilde{\nu}_{\tau}$}
\def\stop{$\tilde{t}_1$ }

\usepackage{color}
\newcommand{\likeJ}{\mathcal{L}_{\rm Joint}}
\newcommand{\like}{\mathcal{L}}
\newcommand{\Ohsq}{\Omega_\chi h^2}
\newcommand{\BR}{BR}
\newcommand\RBtaunu{\frac{\BR(B_u \to \tau \nu)}{\BR(B_u \to \tau \nu)_{SM}}}

\newcommand\brbsmumu{\BR(\overline{B}_s\to\mu^+\mu^-)}
\newcommand\brbdmumu{\BR(\overline{B}_d\to\mu^+\mu^-)}

\newcommand{\brbsgamma}{BR(\bar{B} \rightarrow X_s\gamma) }
\newcommand{\cl}{\text{CL}}

\def\gappeq{\mathrel{\rlap {\raise.5ex\hbox{$>$}}
{\lower.5ex\hbox{$\sim$}}}}
\def\lappeq{\mathrel{\rlap{\raise.5ex\hbox{$<$}}
{\lower.5ex\hbox{$\sim$}}}}

\title{\bf \Large
Supersymmetry Searches in GUT Models \\ with Non-Universal Scalar Masses}

\author[1]{M. Cannoni\thanks{email: mirco.cannoni@dfa.uhu.es}}
\author[2,3]{J. Ellis\thanks{email: John.Ellis@cern.ch}}
\author[1]{M. E. G\'omez\thanks{email: mario.gomez@dfa.uhu.es}}
\author[4]{S. Lola\thanks{email: magda@physics.upatras.gr}}
\author[5]{R. Ruiz de Austri\thanks{email: rruiz@ific.uv.es}}
\affil[1]{\small Departamento de F\'isica Aplicada, Facultad de Ciencias 
Experimentales, \newline  Universidad de Huelva, 21071 Huelva, Spain}
\affil[2]{\small Theoretical Particle Physics and Cosmology Group, Physics
Department, King's College London, \newline London WC2R 2LS, UK}
\affil[3]{\small TH Division, Physics Department, CERN CH-1211 Geneva 23, Switzerland}
\affil[4]{\small Department of Physics, University of Patras, 26500 Patras,
 Greece}
\affil[5]{\small Instituto de F\'isica Corpuscular, IFIC-UV/CSIC, Valencia, Spain}

\date{}
\begin{document}

\maketitle

\begin{center}
~~KCL-PH-TH/2015-52, LCTS/2015-39, CERN-PH-TH/2015-271, \\
UHU-FISUM/2015-17, IFIC/15-89 
\end{center}

\abstract{
We study SO(10), SU(5) and flipped SU(5) GUT models with non-universal soft supersymmetry-breaking
scalar masses, exploring how they are constrained by LHC supersymmetry searches and cold dark
matter experiments, and how they can be probed and distinguished in future experiments. 
We find characteristic differences between the various GUT
scenarios, particularly in the coannihilation region, which is very
sensitive to changes of parameters. For example, the flipped SU(5) GUT
predicts the possibility of $\tilde{t}_1-\chi$ coannihilation, which is absent  
in the regions of the SO(10) and SU(5) GUT parameter spaces that we study.  We use
the relic density predictions in different models to determine upper bounds for 
the neutralino masses, and we find large differences
between different GUT models in the sparticle spectra for the same LSP
mass, leading to direct connections of distinctive possible experimental
measurements with the structure of the GUT group. 
We find that future LHC searches for generic missing $E_T$, charginos and stops 
will be able to constrain the different GUT models in complementary ways, as will the
Xenon 1 ton and Darwin dark matter scattering experiments and
future FERMI or CTA $\gamma$-ray searches.}

\newpage
\tableofcontents

\section{Introduction}
\label{sec:1_Intro}

The recent years have provided a plethora of new experimental and cosmological information 
that provides important constraints on possible extensions of the Standard Model (SM). 
Recent LHC results, including the
Higgs measurements \cite{higgs1,higgs2,MH,muH}, 
severely constrain some of the simplest 
scenarios. We know, however, that there must be some physics beyond the SM.
For example, massive neutrinos
cannot be accommodated within the SM, nor can
the observed baryon asymmetry of the universe or
the origin of Cold Dark Matter (CDM) be explained.
In looking for possible extensions of the SM that address these
issues, supersymmetry (SUSY) continues to provide
significant theoretical advantages, especially if we believe in
unification beyond the SM. Most notably for the purposes of this paper, 
the lightest supersymmetric particle
(LSP) can explain the origin of CDM~\cite{DM-susy_1,DM-susy_2}. 

Reconciling the amount of CDM  deduced from the data of
the Wilkinson Microwave Anisotropy Probe
(WMAP)~\cite{WMAP7,WMAP9} and the Planck satellite~\cite{PLANCK,Ade:2015xua} with the 
predictions of supersymmetric models, has been a major challenge  
in recent years. Although
the minimal constrained supersymmetric extension of the SM (CMSSM) 
is still compatible with the LHC and WMAP
predictions \cite{EO2013}, 
the allowed parameter space is severely constrained,
since a Higgs mass $m_h \sim 125$~GeV implies
a relatively heavy sparticle spectrum.
However, the allowed regions may change significantly in 
different versions of the MSSM, especially 
in the coannihilation strips where, e.g., $m_\chi \sim m_{\tilde{\tau_1}}$  or $\sim m_{\tilde t_1}$. 
Because of their narrow widths,
these coannihilation strips are particularly sensitive to changes in the input model parameters. 

In this work, we study these changes in various scenarios that go
beyond the constrained minimal supersymmetric extension of the SM (CMSSM),
in which soft SUSY-breaking terms are assumed to be universal at the GUT scale, 
using dark matter considerations as a probe
of different theoretical constructions.  In general, GUT scenarios
that favour particular degeneracies in the sparticle spectrum will lead to
additional contributions to coannihilations, thus enhancing their
efficiency.  Conversely, experimental signals sensitive to
these degeneracies can provide information 
about the gauge unification group. Following previous studies of models
with non-universal soft SUSY-breaking Higgs mass
parameters 
(NUHM1,2)~\cite{Ellis_CMSSM_NUHM1,Ellis_NUHM2,Strege:2012bt,Baer:2005bu,Roszkowski:2009sm}, 
we analyse the predictions of various SUSY GUT models, including SO(10)~\cite{SO10_1,SO10_2,SO10_3,SO10_4,SO10_5}, minimal 
SU(5)~\cite{SU5_1,SU5_2} and flipped SU(5)~\cite{FSU5_1,FSU5_2,FSU5_3,FSU5_4}.

Our paper is structured as follows: In Section 2 we review the
features of GUT models that are relevant for our studies. In Section 3
we discuss our sampling methodology for searching for regions of the
parameter space compatible with the data.
In Section 4 we discuss the implications of non-universalities for the
different mechanisms that reproduce the correct relic CDM density.
In Section 5 we discuss in more detail
how the results of our scans depend on the 
relation between the relic abundance mechanisms, the value of the Higgs boson mass
and the supersymmetric contribution to  
$\delta a_\mu=[(g_\mu -2)/2]_\text{exp} -[(g_\mu -2)/2]_\text{SM} $.
In Section 6, we present the cross sections for direct and
indirect CDM detection. In Section 7, we study how sparticle
searches at the LHC impose further constraints on our models. Finally, in Section
8 we summarise our results and discuss future prospects.

\section{Relevant Features of Supersymmetric GUT Models} 
\label{sec:2}

We assume that SUSY breaking occurs at some scale $M_X$ above $M_{GUT}$, 
and is induced by a mechanism that 
generates generation-blind soft terms.  Between the scales $M_X$ and
$M_{GUT}$, the renormalization group equations (RGE) and additional 
interactions associated with flavour, 
e.g., Yukawa interactions,
might induce non-universalities 
in the soft terms, which we do not consider here, while the theory still preserves the GUT symmetry.
Below $M_{GUT}$, the effective theory is the MSSM
with SUSY masses that are common for fields in the same representation of the GUT group. 
Our approach is therefore to assume a pattern of soft terms with common soft masses for all the 
particles that belong to the same representations of the GUT group under consideration, 
while allowing different common masses for inequivalent representations. 

The simplest possibility arises within an SO(10) GUT~\cite{SO10_1,SO10_2,SO10_3,SO10_4,SO10_5}. 
In this case, all quarks and leptons are accommodated in the same
{\bf 16}  representation,  while we assume that the up and down 
Higgs multiplets are in a pair of {\bf 10} representations. 
Since this assignment also determines common sfermion mass
matrices and beta functions, similar behaviour under RGE runs is
to be expected. Consequently, in this GUT model there is a common
soft SUSY-breaking mass for all sfermions (squarks and sleptons)
and two different masses for $m_{h_u}$  and  $m_{h_d}$. 
From this point of view, the SO(10) scenario can be identified with the NUHM2 studied 
previously~\cite{Ellis_NUHM2}, 
and we include it as a reference for comparison with other GUT groups.

The situation changes significantly in the case of the
SU(5) group~\cite{SU5_1,SU5_2}. In this case, the multiplet assignments are as follows:
\begin{equation}
(Q,u^{c},e^{c})_{i} \; \in \mathbf{10}_i \, , \; (L,d^{c})_{i} \; \in \; \mathbf{\overline{5}}_i \, , \; \nu^c_{i} \; \in \; \mathbf{1}_i \, .
\end{equation}
The soft terms that we assume are the same for all the members of the
same representation at the GUT scale, but are different for the $\mathbf{10}$ and 
$\mathbf{\overline{5}}$ in general,
and we assume that the singlet neutrinos decouple at the GUT scale, and therefore do not affect our 
analysis.
A similar approach was followed in~\cite{Okada:2013ija}, 
but the main aim of that work was to reconcile the correct 
prediction of $m_h$ (requiring a heavy  SUSY spectrum) with a supersymmetric
contribution that could explain the discrepancy of the SM prediction for
$(g_\mu -2)$ with its experimental value. 
However, the main aim of this work is to analyse the relic density predictions 
and to extend the full analysis also to the case of the flipped
SU(5)~\cite{FSU5_1,FSU5_2,FSU5_3,FSU5_4}, which leads to predictions that differ significantly, and 
has some distinctive features.
The particle assignments are different in  flipped SU(5)~\cite{FSU5_1,FSU5_2,FSU5_3,FSU5_4}:
\begin{equation}
(Q,d^{c},\nu^{c})_{i} \; \in \mathbf{10}_i \, , \; (L,u^{c})_{i} \; \in \; \mathbf{\overline{5}}_i \, , \; e^c_{i} \; \in \; \mathbf{1}_i \, .
\end{equation}
The impacts of these assignments
on the evolution of sparticle masses, and hence on the coannihilation
strips are discussed in detail below. As before, we assume that
the singlet neutrinos have already decoupled at the GUT scale.
In both SU(5) models we assume that the Higgs doublets $H_u$ and $H_d$
of the MSSM arise from {\bf 5} and $\mathbf{\bar{5}}$ SU(5) representations, respectively.

The soft SUSY-breaking scalar terms for the fields in an irreducible representation $r$ of the 
unification group are parametrised as multiples of a common scale $m_0$: 
\begin{equation}
m_{r}=x_r \, m_{0}, 
\end{equation}
while the trilinear terms are defined as: 
\begin{equation}
A_r = Y_r \,  A_0,  \;\;\; A_0=a_0 \, m_0 \, ,
\end{equation}
where $Y_r$ is the Yukawa coupling associated with the
representation  $r$, and  we use the standard parametrization with $a_0$
a dimensionless factor, which we assume to be representation-independent. 
Since the two Higgs fields of the MSSM arise from different SU(5) representations, they have 
different soft masses, in general. The situation in the different GUT groups is then as follows:

\begin{itemize}
\item SO(10): In addition to the CMSSM parameters, we introduce two new parameters $x_u$ and $x_d$ 
defined as follows: 
\begin{equation}
m_{16}=m_0, \;\;\; m_{H_u}=x_u \, m_{16}, \;\;\; m_{H_d}=x_d \, m_{16} . 
\end{equation}
Similarly, the $A$-terms are parametrised by: 
\begin{equation}
A_{16}=a_0 \cdot m_{0} \, ,
\end{equation}
as in minimal SO(10) with fermion fields in a common {\bf 16} representation 
and two Higgs fields in different {\bf 10} representations. 

\item SU(5): Here we  use as reference the common soft SUSY-breaking masses for the 
fields of the ${\bf 10}$, $m_{10}$. The masses for the other
representations  are then defined as:
\begin{equation}
 m_{10}=m_0, \;\;\; m_5=x_5\cdot m_{10},  \;\;\; m_{H_u}=x_u \cdot m_{10}  \;\;\; m_{H_d}=x_d \cdot m_{10}  \, ,
\end{equation}
and the $A$-terms are specified via a common mass scale:
\begin{equation}
A_{10,5}=a_0 \, m_{0} \, .
\end{equation}
\item Flipped SU(5): Here we have
\begin{equation}
m_{10}=m_0, \;\;\; m_5=x_5\cdot m_{10}  \;\;\; m_R=x_R\cdot m_{10}  \;\;\; 
m_{H_u}=x_u \cdot m_{10}  \;\;\; m_{H_d}=x_d \cdot m_{10}  \, ,
\end{equation}
where $x_R$ refers to the SU(2)-singlet fields.
Similarly, the A-terms are specified as universal: $A_0=a_0\cdot m_0$. 

\end{itemize}
A similar parametrization
of GUT scalar non-universality in SO(10) and SU(5) was used also in Ref.~\cite{Pallis:2003aw}.
For our analysis in the following Sections we assume a common unification scale $M_{GUT}$ defined as 
the meeting point of the $g_1$ and $g_2$ gauge couplings. The GUT value for $g_3$ 
is obtained by requiring $\alpha_s(M_Z)=0.1187$.  Above $M_{GUT}$ we assume a unification 
group that breaks at this scale. We also assume that SUSY is broken above $M_{GUT}$ 
by soft terms that are representation-dependent but generation-blind. 

\section{Sampling Methodology and  Constraints }
\label{sec:3}

We work within the three different GUT scenarios
SO(10), SU(5) and flipped SU(5) described above, mapping the 
areas of the parameter space allowed by WMAP, Planck and other constraints onto the ratios of GUT 
values of the soft terms for each representation. We perform scans of the model parameter spaces using
their matter representation patterns as guides for the soft scalar terms at  the GUT scale,
assuming common gaugino masses. 
For our searches for regions of the parameter spaces compatible with the data, we
use a Bayesian approach based on the {\tt MultiNest} algorithm~\cite{Feroz:2008xx} as implemented in 
{\tt SuperBayeS-v2.0}~\cite{SuperBayes}.
{\tt SuperBayeS} is interfaced with {\tt SoftSUSY} 3.3.10~\cite{SoftSUSY, Allanach:2001kg} 
as SUSY spectrum calculator, {\tt MicrOMEGAs} 2.4~\cite{MicrOMEGAs, Belanger:2006is} 
to compute the abundance of dark 
matter, {\tt DarkSUSY} 5.0.5~\cite{DarkSUSY, Gondolo:2004sc} for the computation of $\sigma_{SI}$, 
{\tt SuperIso} 3.0~\cite{SuperIso, Mahmoudi:2008tp} to compute $\delta a_\mu^{\mathrm{SUSY}}$ 
and $B(D)$ physics observables, {\tt SusyBSG 1.5} for the determination of 
$\brbsgamma$\cite{SusyBSG,Degrassi:2007kj}.

The likelihood function that drives our exploration to regions of the parameter space where 
the model predictions fit the data well is built from the following components:
\begin{equation}
\begin{aligned}
\ln \likeJ &= \ln \like_{\rm EW}+ \ln \like_{\rm B} + \ln \like_{\Ohsq} \\
&+ \ln \like_{\rm LUX} + \ln \like_{\rm Higgs} + \ln \like_{\rm SUSY} + \ln \like_{\rm g-2},
\end{aligned}
\label{eq:like}
\end{equation}
where $\like_\text{EW}$ is the part corresponding to electroweak precision observables,
$\like_\text{B}$ to B-physics constraints, $\like_{\Ohsq}$ to measurements of the cosmological DM 
relic density, $\like_\text{LUX}$ to the constraints from
direct DM detection searches (dominated by the LUX experiment) and $\like_\text{Higgs}$
($\like_\text{SUSY}$) to Higgs (sparticle) searches at colliders.
We now discuss each component in turn:

\paragraph{$\like_{\rm EW}$:} 
We implement the constraints on the effective electroweak mixing 
angle  $\sin^2\theta_\text{eff}$ and the total width of the Z-boson, $\Gamma_{Z}$, from 
the LEP experiments~\cite{ALEPH:2005ab}.
For the mass of the W boson, $m_W$, we use the 
Particle Data Group value~\cite{PDG}, which combines the LEP2 and Tevatron measurements.
We assume Gaussian likelihoods for all these quantities, with means and standard deviations 
as given in Table~II of \cite{Strege:2014ija}.

\paragraph{$\like_{\rm B}$:} We consider the following flavor observables related to B physics:
$\brbsgamma$, $R_{\Delta M_{B_s}}$, $\RBtaunu$,
$\brbsmumu$ and $\brbdmumu$. We assume Gaussian likelihoods for all of them, and for most of them we 
use the measurements shown in Table II of \cite{Strege:2014ija}. However, the experimental values
assumed for $\brbsmumu$ and $\brbdmumu$ are $(2.9 \pm 0.8) \times 10^{-9}$ and $(3.6 \pm 1.55) 
\times 10^{-10}$, respectively, where we quote the total uncertainties found by adding
in quadrature the theoretical~\cite{Arbey:2012ax} and experimental~\cite{CMSandLHCbCollaborations:2013pla,CMS:2014xfa} uncertainties.

\paragraph{$\like_{\Ohsq}$:} We include the constraint 
on the DM relic abundance from the Planck satellite, assuming that the lightest neutralino
is the dominant DM component. We use as central value the result from Planck temperature and 
lensing data $\Ohsq = 0.1186 \pm 0.0031$ \cite{Ade:2013zuv}, with a (fixed) theoretical 
uncertainty 
$\tau = 0.012$,  following Refs. \cite{Strege:2012bt,Roszkowski:2009sm,Roszkowski:2014wqa}, to account for 
the numerical uncertainties entering in the calculation of the relic density.

\paragraph{$\like_{\rm LUX}$:} For direct DM detection, we include the upper limit from the LUX 
experiment \cite{LUX_PRL}, as implemented in the {\tt LUXCalc}
code~\cite{Savage:2015xta}, including both the spin-independent and spin-dependent 
cross-sections in the event rate calculation. We adopt
hadronic matrix elements determined by lattice QCD \cite{QCDSF:2011aa,Junnarkar:2013ac}.

\paragraph{$\like_{\rm Higgs}$:} The likelihood for the Higgs searches has two components. The 
first implements bounds obtained from Higgs searches at LEP, Tevatron and LHC via
{\tt HiggsBounds} \cite{Bechtle:2013wla}, which returns whether a model is excluded or not at the 
95\% \cl.
The second component constrains the mass and the production times decay rates of the 
Higgs-like boson discovered by the LHC experiments
ATLAS \cite{higgs1} and CMS \cite{higgs2}. For this we use {\tt HiggsSignals} 
\cite{Bechtle:2013xfa}, assuming a theoretical uncertainty in the calculation of the lightest Higgs mass 
of 2 GeV.

\paragraph{$\like_{\rm SUSY}$:} The constraints from SUSY searches at LEP and Tevatron are evaluated following the 
prescription proposed in~\cite{deAustri:2006pe}. The present limits from the Run 1 of LHC are displayed in the corresponding Figures in Section 4.  

\paragraph{$\like_{\rm g-2}$:} We adopt for the discrepancy between the experimental value of the anomalous magnetic moment of the muon 
and the value calculated in the Standard Model $\delta 
a_{\mu}^{\text{SUSY}} = (28.7 \pm 8.2) \times 10^{-9}$ \cite{Davier:2010nc}, where 
experimental and theoretical errors have been added in quadrature. This corresponds to a 
$3.6\sigma$ discrepancy with the value predicted in the Standard Model, and relies
on $e^+e^-$ data for the computation of the hadronic loop contributions to the Standard Model 
value. The likelihood function is assumed to be Gaussian.

~~\\
In each case, we run the {\tt MultiNest} algorithm until we reach a sample of about $3\times 10^4$ points. Our focus in this work is to scan the parameter space
of the new models, in order to study their phenomenology and
identify regions compatible with the data. Performing 
any statistical (frequentist or Bayesian) 
interpretation of our results, based on global fits and confidence or
credibility level regions, is beyond the scope of this paper and
would require samples that are
orders of magnitude larger that the ones  we have gathered.
Instead, we present scatter plots showing the correlations of pairs of parameters and/or observables in various  planes.
In doing this, we select from the full samples only those
points  predicting the value of 
all the observables within the 2$\sigma$ interval 
(with $\sigma$ obtained by summing in quadrature 
the experimental and theoretical errors as explained in the previous
paragraphs). 
If for the observable
only an experimental exclusion limit exists, then the theoretical value is required to be within 
the 90/95 \% \cl\ exclusion limits.
After applying these cuts, the number of points in the samples is substantially reduced.
In particular, in none of the GUT models do we find points
with a supersymmetric contribution to $\delta a_\mu$ within the
$2\sigma$ interval. However, we highlight the points in our samples whose contributions 
to the anomalous magnetic moment of the muon lie in the $3\sigma$ interval.
We discuss this issue in Section~\ref{sec:5}.  

\section{ Non-Universality Parameters and Relic Density \\Mechanisms}
\label{sec:4}

It is well known that, if the required 
amount of relic dark matter is provided by neutralinos,
then particular mass relations must be present in the 
supersymmetric spectrum. In addition to mass relations, we use the neutralino composition
to classify the relevant points of the supersymmetric parameter space.
The higgsino fraction of the lightest neutralino mass eigenstate is characterized by
the quantity
\begin{equation}
h_f \; \equiv \; |N_{13}|^2 + |N_{14}|^2 \, ,
\end{equation}
where the $N_{ij}$ are the elements of the unitary 
mixing matrix that correspond to the higgsino mass states.
Thus, we classify the points that pass the constraints discussed in Section 2 according 
to the following criteria:
\paragraph{Higgsino \LSP  :}
\begin{flalign}
h_f >0.1, \;\;|m_A-2 m_\chi| > 0.1 \, m_\chi.
\label{criterio_higgsino}
\end{flalign}
In this case, the lightest neutralino is  higgsino-like and, as we discuss later, 
the lightest chargino $\chi^\pm_1$ is almost degenerate in mass with \LSP.
The couplings to the SM gauge bosons are not suppressed and \LSP\ pairs have large 
cross sections for annihilation into $W^+ W^-$ and $ZZ$ pairs, which may reproduce the observed value
of the relic abundance. Clearly, coannihilation channels involving $\chi^\pm_1$ and 
$\chi^0_2$ also contribute. 

\paragraph{$A/H$ resonances:}
\begin{flalign}
|m_A-2 m_\chi|\leq 0.1 \, m_\chi.
\label{criterio_res}
\end{flalign}
The correct value
of the relic abundance is achieved thanks to $s$-channel annihilation,
enhanced by the resonant $A$ propagator. The thermal average $\langle \sigma_{ann}v\rangle$
spreads out the  peak in the cross section, so that neutralino masses for 
which $2m_{\chi} \simeq m_A$ is not exactly realized can also experience resonant annihilations.
\paragraph{$\tilde{\tau}$ coannihilations:}
\begin{flalign}
h_f <0.1,\;\;(m_{\tilde{\tau}_1}-m_\chi)\leq 0.1 \, m_\chi
\label{criterio_RR}
\end{flalign}
The neutralino is bino-like, annihilation into leptons through $t$-channel slepton exchange 
is suppressed, and coannihilations involving the nearly-degenerate \stau \ are necessary 
to enhance the thermal-averaged effective cross section.
\paragraph{$\tilde{\tau}-\tilde{\nu}_\tau$ coannihilations:}
\begin{flalign}
h_f <0.1,\;\;(m_{\tilde{\tau}_1}-m_\chi)\leq 0.1 \, m_\chi,\;\;(m_{\tilde{\nu}_\tau}-m_\chi) \leq  0.1 \, m_\chi.
\label{criterio_LL}
\end{flalign}
Similar to the previous case, but also the $\nu_{\tilde \tau}$ is nearly degenerate in mass with the 
\stau.
\paragraph{$\tilde{t_1}$ coannihilations:}
\begin{flalign}
h_f <0.15,\;\;(m_{\tilde{t}_1}-m_\chi)\leq 0.1 \, m_\chi.
\label{criterio_stop}
\end{flalign}
The \stop \ is light and nearly degenerate with the bino-like neutralino.
These coannihilations are present in the flipped SU(5) model.

We have performed the parameter-space scans in the three GUT groups with two
different sets of ranges, as detailed in Table~\ref{tab1}.
The first one (Set 1) is broader,
sampling soft terms up to 10 TeV and all the $x_i$ in the ranges $0< x_i <2$. 
The {\tt MultiNest} sampling of Set 1 finds that the data
are  more easily accommodated with a heavy spectrum, where  the higgsino neutralino 
and $A$ funnel mechanisms dominate and only few points in the coaannihilation areas are found
within the $3\times 10^4$ points sample. 

Therefore, in order to zoom in the low mass spectrum, 
where coannihilations are expected to show up and is also favoured
by the $\delta a_\mu$ constraint, 
we performed  a separate scan (Set 2), 
where we decrease the upper limits on
$m_0$ and $m_{1/2}$.
Furthemore, the ranges of the paramters $x_i$ are also restricted, since 
the coannihilation regions depend
on the values of the $x_i$ in a known way, as we explain below.

\begin{table}[t!]
\begin{center}
\begin{tabular}{cccc}
\hline  
\hline
\rule[-2ex]{0pt}{5.ex} \bf{Set 1} & SO(10)  & SU(5)  & FSU(5)
 \\ 
\rule[-2ex]{0pt}{3.ex} $100\;\text{GeV}\leq  m_0 \leq 10\; \text{TeV}$  & $0\leq x_u \leq 2$ & $0\leq x_u \leq 2$ & $0\leq x_u \leq 2$ 
\\ 
\rule[-2ex]{0pt}{3.ex} $50\; \text{GeV}\leq  m_{1/2} \leq 10\; \text{TeV}$ & $0\leq x_d \leq 2$ & $0\leq x_d \leq 2$ & $0\leq x_d \leq 2$ 
\\ 
\rule[-2ex]{0pt}{3.ex} $-10\;  \text{TeV}\leq  A_{0} \leq 10\; \text{TeV}$  &  & $0\leq x_5 \leq 2$ & $0\leq x_5 \leq 2$ 
\\ 
\rule[-2ex]{0pt}{3.ex} $2\leq \tan\beta \leq 65$ &  &  &  $0\leq x_R \leq 2$ 
\\ 
\hline
\hline 
\rule[-2ex]{0pt}{5.ex} \bf{Set 2}   & SO(10) &  SU(5)&  FSU(5)
\\ 
\rule[-2ex]{0pt}{3.ex} $100\; \text{GeV}\leq m_0  \leq 2500\;\text{GeV}$   & $0\leq x_u \leq 1$ & $0\leq x_u \leq 1$ &  $0\leq x_u \leq 1$
\\ 
\rule[-2ex]{0pt}{3.ex} $50\; \text{GeV}\leq  m_{1/2} \leq 2500\; \text{GeV}$  & $0\leq x_d \leq 2$ & $0\leq x_d \leq 2$ &  $0\leq x_d \leq 2$
\\ 
\rule[-2ex]{0pt}{3.ex} $-10\;  \text{TeV}\leq A_{0} \leq 10\; \text{TeV}$  &  & $0\leq x_5 \leq 2$  &  $0\leq x_5 \leq 1$
\\ 
\rule[-2ex]{0pt}{3.ex} $2\leq  \tan\beta  \leq 65 $ &  &  & $1\leq x_R \leq 2$  \\ 
\hline \hline
\end{tabular}
\caption{\it Sets of ranges used to sample the parameter spaces of the GUT models defined 
in Section~\ref{sec:2}. See the text for the definitions of the non-universality parameters $x_i$.   }
\label{tab1}
\end{center}
\end{table}

\begin{figure}[t!]
\begin{center}
\includegraphics*[scale=0.5]{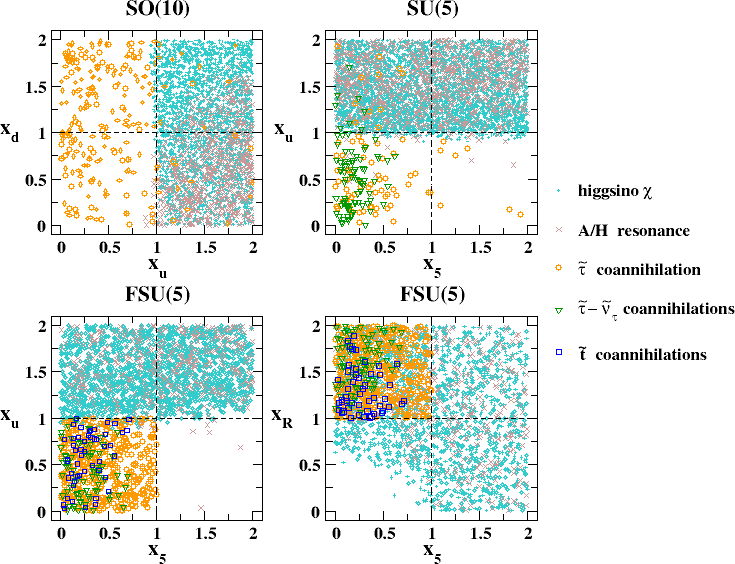}
\caption{Correlations between the non-universal soft SUSY-breaking scalar mass parameters and the
relic density mechanisms. 
The legend showing the meanings of colours and symbols for the points applies to 
all the figures of this paper.  
}
\label{fig:GUTpar}
\end{center}
\end{figure}

In the following plots, the points corresponding to the above mechanisms will 
be presented using different symbols and colours, as specified in 
the legend of Fig.~\ref{fig:GUTpar}.
Although, in the scatter plots that we present, points of different types appear superimposed, 
we found that by applying the selection rules (\ref{criterio_higgsino})-(\ref{criterio_stop}) 
to the whole samples, the obtained sets do not intersect. 
Only a few points that have $h_f >0.1$ and do not satisfy any of the 
above conditions are found in the Set~1 scans. These points lie 
in the areas of both the $A/H$ resonances 
and a higgsino-like neutralino. Since they do not form a distinct
region, they will not be shown in the figures, for reasons of clarity.

Fig.~\ref{fig:GUTpar} shows the correlations between the non-universal parameters and 
the relic density mechanisms for each GUT group.
The parameter $x_d$ has no particular correlation with the above mechanisms,
while in all cases higgsino DM corresponds to $x_u>1$, and the same is also true
for almost all the $A/H$ funnel points, as seen in the upper left, upper right and lower left panels of Fig.~\ref{fig:GUTpar}.
These mechanisms are  independent from the GUT model choice, because the Higgs mass parameters  
follow the same pattern in the three scenarios. They are also influenced 
by the other soft terms due to renormalization group running, but this effect is not  
apparent in scatter plots like those presented here. 

In SO(10) and SU(5) most of the $\tilde{\tau}$ coannihilation points 
have $x_u <1$. 
In SU(5), the flexibility allowed by the $x_5$ parameter allows
$\tilde{\tau}-\tilde{\nu}$ coannihilations that are not present in SO(10). As seen in Fig.~\ref{fig:GUTpar}, these points 
lie in the $x_5 <1$ region. This is due to the fact that the lepton doublet 
belongs to the {\bf 5} representation  while the quarks and the lepton singlet are in the {\bf 10}. 
Then, to satisfy the $m_h$ constraint, the squark masses have to be large 
and the same is true for the right sleptons. However, 
left-slepton soft masses driven by a small value of $x_5$ may lead to sleptons in
the coannihilation range.
The lower panels of Fig.~\ref{fig:GUTpar} show that 
in the flipped SU(5) $\tilde{\tau}$ and $\tilde{\tau}-\tilde{\nu}$ coannihilations
are located in the quadrants defined by $x_u<1$, $x_5<1$, $x_R >1$.
We also see that a $\tilde{t}$ coannihilation area is present.
This is possible in flipped SU(5) because the right-handed squarks are in the $\mathbf{\overline{5}}$ representation, 
so the stop mass decreases with $x_5$. On the other hand,
$x_R$  cannot be very small or the lightest stau becomes tachyonic. 
By restricting it to be $> 1$ we 
avoid this situation, and also increase the left component in the 
lighter stau, since $x_5<x_R$. 

For simplicity and to avoid any discussion about cosmological constraints on tachyons, we consider 
here only positive values for  $m^2_{H_u}$  and $m^2_{H_d} $ at the GUT scale. 
However, some negative values may be allowed, and have been included in some other analyses, leading 
to small differences. 
For example, in the case of the NUHM2 with
$m^2_{H_u} \neq m^2_{H_d}\neq m^2_0$,
the authors of~\cite{Ellis_NUHM2} also consider negative values for 
$m_0$, $m^2_0$, $m^2_{H_u}$, $m^2_{H_d}$ and find
their best fit point for $m_0 <0$. In \cite{Bagnaschi:2015eha} 
the authors find a small stop island at 95\% CL in the NUHM1 ($m^2_{H_u} =
m^2_{H_d}\neq m^2_0$) and  a larger
one in the NUHM2; this can be attributed to negative values of
$m^2_{H_u} $, which enters in the RGE for $m_{\tilde{t}_R}^2$
and decreases its value. 
Because of our restriction to positive values of $x_u$ and $x_d$, 
in our analysis flipped SU(5) is the only scenario where this mass can become low for low values of $x_5$, due to its presence in the $\mathbf{\overline{5}}$
instead of the $\mathbf{10}$.
In~\cite{Okada:2013ija}, $M_A$ and $\mu$ are taken as free
parameters at low energies and  the RGEs
are used to obtain the corresponding  GUT values for
$m^2_{H_u}$ and  $m^2_{H_d}$, as described in~\cite{Baer:2005bu}. This is
a way to  avoid sampling points that fail the electroweak symmetry breaking  test. 
However, although some of these points correspond to negative values for  
$m^2_{H_u}$ and/or $m^2_{H_d}$, the
authors do not display any region with  stop coannihilations.

\section{Relic Density, Higgs Mass and  $\delta a^{SUSY}_\mu $}
\label{sec:5}
\begin{figure}[t!]
\begin{center}
\includegraphics*[scale=.5]{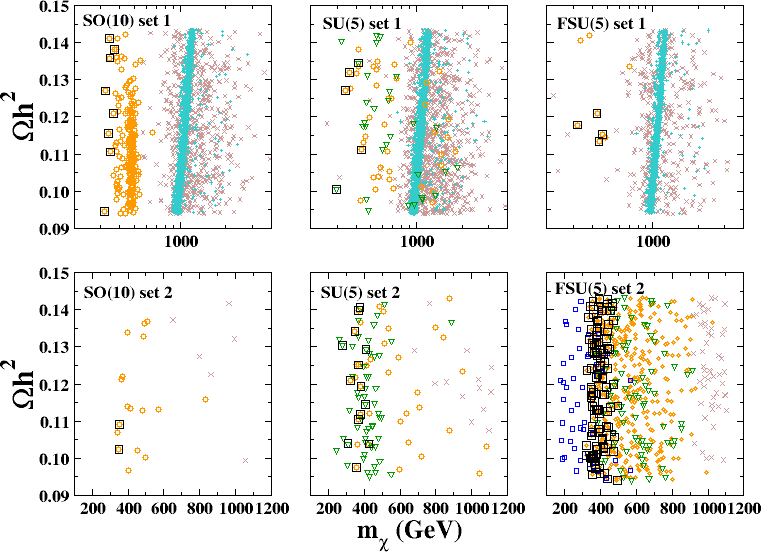}
\caption{
The neutralino relic density $\Ohsq$ as a function of the neutralino mass, using the symbols
defined in the legend of Fig.~\ref{fig:GUTpar}.
The points surrounded by black squares satisfy the constraint $\delta a_\mu$ at 3$\sigma$.}
\label{omegah2}
\end{center}
\end{figure}
We now discuss in more detail the relations found in our scans
between the relic abundance mechanisms, the value of the Higgs boson mass
and the supersymmetric contribution to  
$\delta a_\mu=[(g_\mu -2)/2]_\text{exp} -[(g_\mu -2)/2]_\text{SM} $.

Most of the turquoise points in Fig.~\ref{omegah2},
with a higgsino \LSP\ are confined in a thin strip with mass around 1 TeV, independently 
of the gauge group,
and are only present in the upper panels of Fig.~\ref{fig:GUTpar} where $x_u >1$, as discussed 
previously. A  higgsino-like neutralino with mass around 1 TeV is a general prediction
driven by the relic density bound and has been emphasized 
before in many analysis~\cite{Olive:1990qm}, \cite{Profumo:2004at}, \cite{Chattopadhyay:2005mv}, 
\cite{Roszkowski:2014wqa}, \cite{Olive:2015maa}.

Most of the $A/H$ resonance points have a \LSP\ mass larger than 800-900 GeV. They are 
numerous in the Set 1 scans (upper panels of Fig.~\ref{omegah2}), 
whereas they are reduced substantially in the Set 2 scans (lower panels of Fig.~\ref{omegah2}). 
In fact, for parameters within the ranges
of Set 2, the $A/H$ mass is smaller than in Set 1, therefore  its decay width is smaller 
and the condition (\ref{criterio_res}) is more difficult to respect.

The coannihilation areas are different in the various models and, as is well known, they 
feature upper limits on the \LSP\ mass.
In the case of SO(10), the \stau\ area (orange circles) is well defined, with the neutralino mass in 
the approximate interval 300-600 GeV.
In the case of SU(5), \staunu\ coannihilations (green triangles) are also involved, and the upper 
limit increases to $\sim 1.1$~TeV.
The number of \stau\ points is reduced drastically in the Set 1 scan of flipped SU(5) 
and in the Set 2 scan of SO(10). In the former case, the right-handed slepton mass is determined by 
the 
parameter $x_R$, which in Set 1 is free to vary over values larger than 1. In the latter case, the reduction is due to tension between 
the contribution to $\delta a_\mu$, which needs relatively light sleptons and gauginos,
and a Higgs mass around 125 GeV, which pushes the preferred values of $m_0$ and $m_{1/2}$
towards higher values. The coannihilations 
in flipped SU(5) are recovered in the scan with Set 2, as seen in the bottom-right panel,
where we also note the appearance of the \stop\ strip (dark blue squares). 
In SU(5) the situation is intermediate. The number of coannihilations does not vary so 
strongly, but in passing from Set 1 to Set 2 a greater concentration of points with 
light neutralino masses can be observed.

The black squares highlight the points for which the supersymmetric contribution
to $\delta a_\mu$ differs from the central value by less than 3$\sigma$. The typical values
of $\delta a_\mu$ in the black squares are found to be in the range $4-6 \times 
10^{-10}$, which is similar to the best-fit points in NUHM1 and NUHM2 
models~\cite{Ellis_CMSSM_NUHM1,Ellis_NUHM2}. 
All the black squares are in the region of \stau\ coannihilation, with a minority
also featuring \staunu\ coannihilations. The scan with the largest number of
$\delta a_\mu$-friendly points occurs in flipped SU(5) Set 2, due to
the lightness of the spectrum and the additional freedom in the choice
of parameters, as compared to SO(10) and SU(5).

In all our models, despite the larger freedom in the scalar sector allowed by the new parameters,
it is hard to fully  explain the $\delta a_\mu$ anomaly with a supersymmetric contribution.
In this respect the situation is thus similar to other models with gaugino and sfermion mass 
unification such as the CMSSM, NUHM1 and NUHM2 models~\cite{Ellis_CMSSM_NUHM1,Ellis_NUHM2}.
The difficulty to explain the anomaly at the 2$\sigma$ level in non-universal scalar GUT models
was also recently discussed in~\cite{Chakrabortty:2015ika}, while,
by relaxing also the condition of gaugino universality, the authors 
of~\cite{Kowalska:2015zja} find models with supersymmetric contribution at $2\sigma$.
Anyway, the GUT-boundary conditions employed in both these studies are different from ours.
\begin{figure}[t!]
\begin{center}
\includegraphics*[scale=0.5]{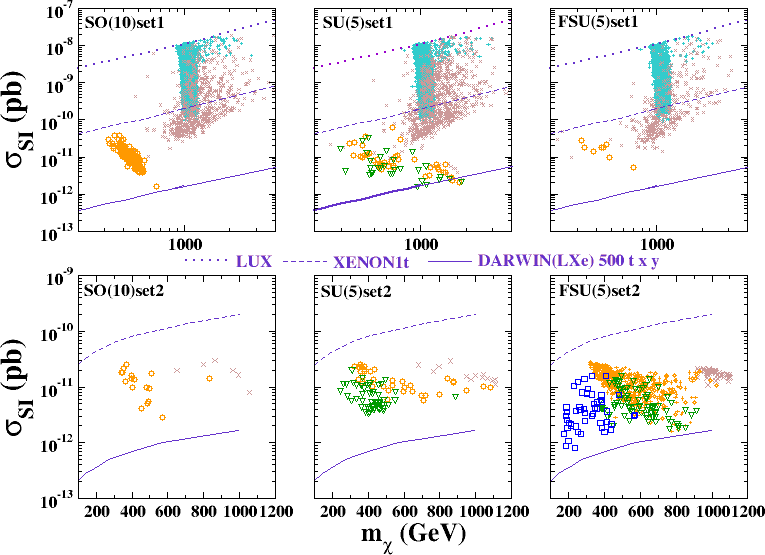}
\end{center}
\caption{The spin-independent neutralino-nucleon cross section as a function of the 
neutralino mass. The dotted line is the current  exclusion curve from the LUX experiment~\cite{LUX_PRL}.
The projected sensitivities at 90\% confidence level for the XENON 1 ton experiment~\cite{XENON1Tproc,Aprile:2015uzo} (dashed line)
and the DARWIN experiment~\cite{DARWINproc} (full line) are taken from~\cite{Xe_sensitivity}. See the legend.  }
\label{fig:sigmaSI}
\end{figure}
\begin{figure}[tb!]
\begin{center}
\includegraphics*[scale=.5]{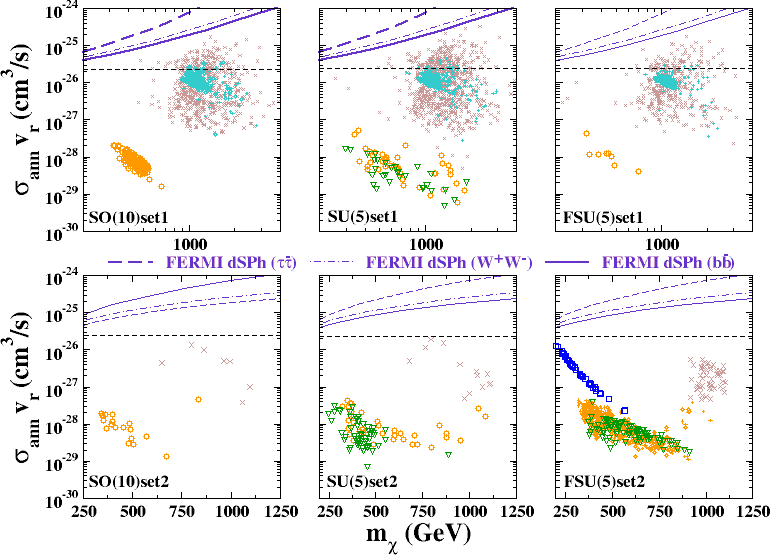}
\end{center}
\caption{The total non-relativistic \LSP\  annihilation cross section times 
relative velocity as a function of the neutralino mass. The purple lines are the present 
exclusion limits from a FERMI analysis of gamma-ray emission from dwarf spheroidal galaxies~\cite{FERMI_dwarfs}:
see the legend for the specific final states. The horizontal black dashed line corresponds to the usual benchmark value of 
$\langle \sigma_{eff} v_\text{rel} \rangle \simeq 2-3 \times 10^{-26}$ cm$^3$/s.}
\label{fig:sigmaann}
\end{figure}

\section{Direct and Indirect Dark Matter Searches}

In Figure \ref{fig:sigmaSI} we show scatter plots of the spin-independent neutralino-nucleon cross section as a 
function of the neutralino mass. The present limits from the null result of the LUX 
experiment~\cite{LUX_PRL} (dotted line)  already exclude points with
a higgsino-like neutralino and 
some points in the $A/H$ funnel area. The projected sensitivity of the XENON 1 ton 
experiment~\cite{XENON1Tproc,Aprile:2015uzo} shows that it could probe most of these areas, 
while coannihilations could be fully probed only 
with a multi-ton mass experiment like the DARWIN project~\cite{DARWINproc}, 
with an exposure of 500 $t\times y $. 
These sensitivity curves are deduced from the recent study in~\cite{Xe_sensitivity}.

We also show in Fig.~\ref{fig:sigmaann} the present situation of the indirect dark matter
search through $\gamma$-ray emission from annihilations in the halos of dwarf galaxies 
of the local group~\cite{SanchezConde:2011ap}, 
by showing the total non-relativistic neutralino-neutralino annihilation 
cross section times the relative velocity in dark matter halos $\sigma_\text{ann} v_r$ as a 
function of the neutralino mass.
The three curves are all limits from the combined analysis of the FERMI satellite with 6 
years of data~\cite{FERMI_dwarfs} obtained assuming that $\tau\bar{\tau}$,
$b\bar{b}$ and $WW$ final states dominate.
Gamma rays may result from the  decays and hadronization of any of these final states, and
in principle these limits apply only to points where these channels dominate 
neutralino annihilation, and can therefore be compared with the higgsino and resonance regions.
We see that at present the curves do not touch the favoured regions of parameter space. 
Future data from FERMI or CTA arrays~\cite{Roszkowski:2014wqa}, \cite{Carr:2015hta} 
may possibly probe the turquoise and red 
points (higgsino and $A/H$ funnel regions). 
As could be expected, the annihilation cross section in the slepton coannihilation region
is too small to be probed by this kind of indirect searches.

The dashed line corresponds to the usual benchmark value of 
$\langle \sigma_{eff} v_\text{rel} \rangle \simeq 2-3 \times 10^{-26}$ cm$^3$/s
for a weakly-interacting massive particle with a relic abundance  $\Omega h^2\simeq 0.1$.
We remark that the values of $\sigma_\text{ann} v_r$ shown in Figure \ref{fig:sigmaann}
coincide with those of the thermal average at freeze-out $\langle \sigma_{eff} v_\text{rel} \rangle$
only when there are no coannihilation channels and the product $\sigma_\text{ann} v_r$ is a constant
independent from the relative velocity/temperature at freeze-out.

\section{LHC Searches}

\begin{figure}[t!]
\begin{center}
\includegraphics*[scale=.5]{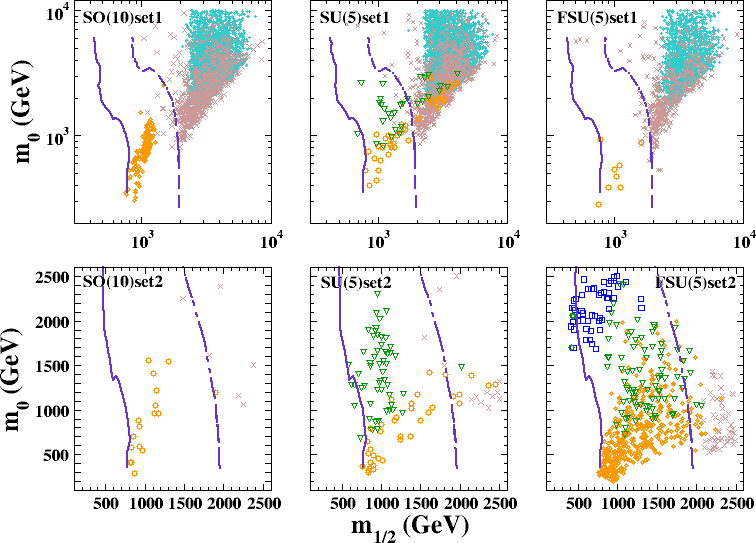}
\end{center}
\caption{Scatter plots of non-universal GUT models in the ($m_{1/2}, m_0$) plane, with the same 
legend as in Fig.~\ref{fig:GUTpar}. The current LHC 95 \% CL exclusion (solid purple line)
and the projected exclusion sensitivity at 14 TeV with 300 fb$^{-1}$ (dashed purple line)
are taken from Ref.~\cite{Bagnaschi:2015eha}.}
\label{fig:mhf_m0}
\end{figure}

\begin{figure}[htbp!]
\begin{center}
\includegraphics*[scale=.5]{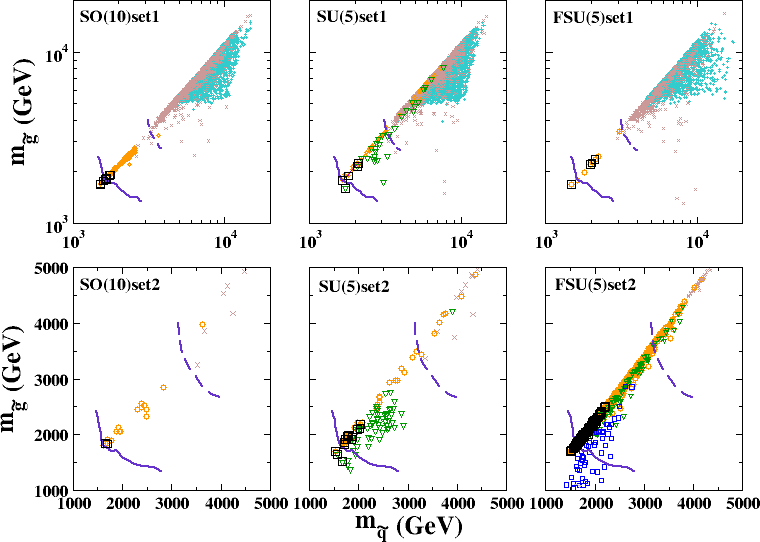}
\end{center}
\caption{Scatter plots of non-universal GUT models in the ($m_{\tilde{q}}, m_{\tilde{g}}$) plane, with the
same legend as in Fig.~\ref{fig:GUTpar}. The current LHC 95 \% CL exclusion (solid purple line)
and the projected exclusion sensitivity at 14 TeV with 300 fb$^{-1}$ (dashed purple line)
are taken from Ref.~\cite{Bagnaschi:2015eha}.}
\label{fig:mg_mq}
\end{figure}

\paragraph{LHC missing energy searches:}
The coannihilation areas yield a light supersymmetric spectrum that
is already partially probed by the first years of operation of LHC.
In Figs.~\ref{fig:mhf_m0} and \ref{fig:mg_mq}, 
we show the distributions in the ($m_{1/2}, m_0$)  and ($m_{\tilde{q}}, m_{\tilde{g}}$) 
planes for the various GUT models. 
The present LHC 95 \% CL exclusion limit from missing $E_T$ searches~\cite{Bagnaschi:2015eha} are
depicted as solid lines, while the projected sensitivity with 300 fb$^{-1}$ at 14 TeV~\cite{Bagnaschi:2015eha}
is represented by dashed lines.
We see that the present exclusion limits already constrain the $\tilde{t}$-coannihilation area
of flipped SU(5) (blue squares), and graze the $\tilde{\tau}$-coannihilation points
favoured by the $\delta a_\mu$ constraint in all GUT models.
The projected LHC missing $E_T$ sensitivity covers most of the coannihilation areas, but 
leaves practically untouched the higgsino and resonance areas.

\paragraph{Heavy Higgs and charginos:}
\begin{figure}[]
\begin{center}
\includegraphics*[scale=.5]{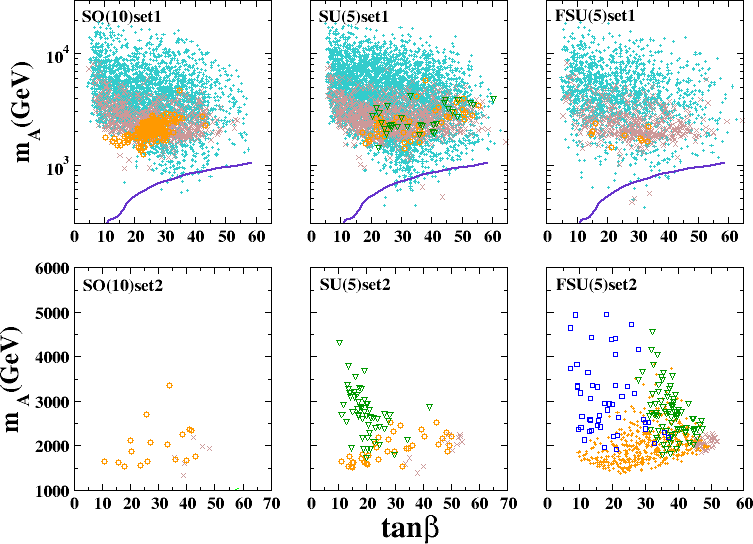}
\end{center}
\caption{Scatter plots of non-universal GUT models in the ($\tan\beta$, $m_A$) plane, with the
same legend as in Fig.~\ref{fig:GUTpar}. The current LHC 95 \% CL exclusion
(solid purple line) is taken
from~\cite{Bagnaschi:2015eha}.}
\label{fig:mA_tb}
\end{figure}
\begin{figure}[htbp!]
\begin{center}
\includegraphics*[scale=.46]{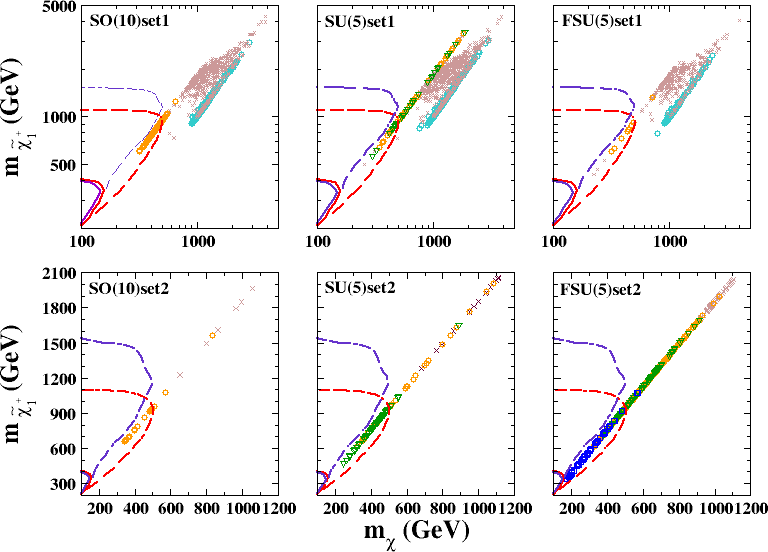}
\end{center}
\caption{Scatter plots of non-universal GUT models in the ($m_\chi $, $m_{\chi^{\pm}_1}$) plane, with the
same legend as in Fig.~\ref{fig:GUTpar}. The current LHC 95 \% CL exclusion (solid purple line) and projected
sensitivity (dashed purple line) are taken from~\cite{Bagnaschi:2015eha}. See the text for details. 
The projected lines correspond to the sensitivity with 3000 fb$^{-1}$.
}
\label{fig:mch1_mlsp}
\end{figure}
We now discuss the sensitivity of other search channels  at the LHC to supersymmetric particles in the models we study.
The present exclusion curve in the ($\tan\beta, m_A$) plane
is shown as a solid purple line in Fig.~\ref{fig:mA_tb}.
We see that the mass of the pseudoscalar neutral Higgs $A$ is generally larger than 1 TeV,
except for a few points in the higgsino and resonance regions, which are excluded by this constraint.
If the sensitivity in this plane could be pushed to masses up to 2-2.5 TeV,
most of the \stau\ coannihilation areas could be probed, as seen in the lower panels of Fig.~\ref{fig:mA_tb}.

More interesting is the search for the lighter chargino,
$\chi^{\pm}_1$, shown in the ($m_\chi $, $m_{\chi^{\pm}_1}$) plane in Fig.~\ref{fig:mch1_mlsp}.
In the Set 1 scans (upper plots), the points are distributed in the region where 
$m_\chi \lesssim m_{\chi^{\pm}_1} \lesssim 2 m_\chi$. 
In the coannihilation regions, the neutralino is bino-like, $m_\chi \simeq M_1$, whereas
the chargino is gaugino-like with $m_{\chi^{\pm}_1}\simeq M_2\simeq 2 M_1\simeq 2M_1$. 
On the other hand, for a higgsino-like neutralino, we have $m_\chi
\simeq \mu$, since 
the mass of the lightest chargino is dominated by the $\mu$ mass parameter.
The $A/H$ funnel resonance, with bino-higgsino neutralino mixing,
lies in the area between the two extreme cases above. 
This behaviour results from the interplay between the universality 
of the gaugino masses at the GUT scale
and constraints imposed by the relic abundance.

The solid indigo line is the present limit from the search for
$\chi^{\pm}_1 \chi^0_2$ production
and decays into  $W/Z$ and missing $E_T$, and 
the solid red line shows the limit from
the search for gaugino pair production  $\chi^{\pm}_1 \chi^0_2, \chi^{\pm}_1 \chi^{\pm}_1$, with multi-$\tau$ 
final-state decays and missing energy. In both cases, the dashed lines indicate the projected sensitivity
with 3000 fb$^{-1}$ at 14 TeV: all the limits are taken from Refs.~\cite{ATLAS_charginos},
\cite{Bagnaschi:2015eha}.
In the latter channel, the LHC will be able to probe parts of the \stau\ coannihilation 
strips (orange circles and green triangles) and, in the case of the flipped SU(5), most of the \stop\ coannihilation strip.

\paragraph{Third-Generation Squarks:}

\begin{figure*}[htbp!]
\begin{center}
\includegraphics*[scale=.46]{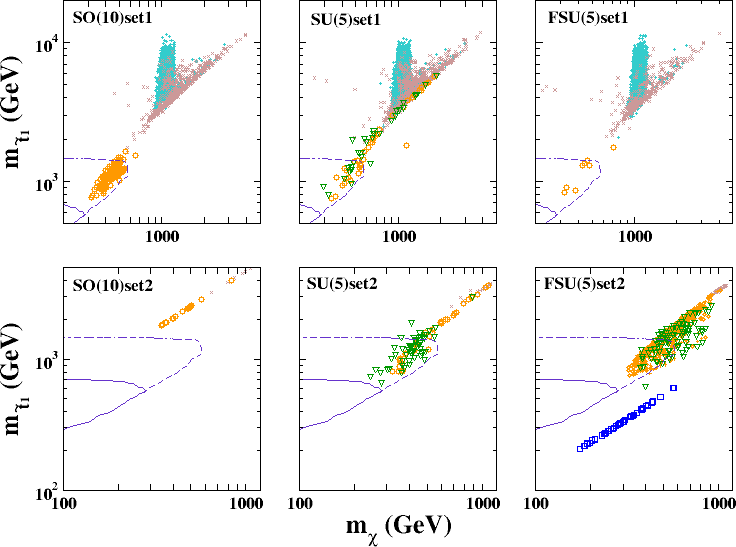}
\end{center}
\caption{ Scatter plots of non-universal GUT models in the ($m_\chi $, $m_{\tilde{t}_1}$) plane, with the same
legend as in Fig.~\ref{fig:GUTpar}. The solid and dashed purple lines  are the present limit and projected
sensitivity
\cite{ATLAS_3rd_squarks_1,Aad:2013ija_3rd_squarks_2,Bagnaschi:2015eha}. See
the text for details. The projected line corresponds to the sensitivity with 3000 fb$^{-1}$.
}
\label{fig:mst1_mlsp}
\end{figure*}
\begin{figure*}[]
\begin{center}
\includegraphics*[scale=.48]{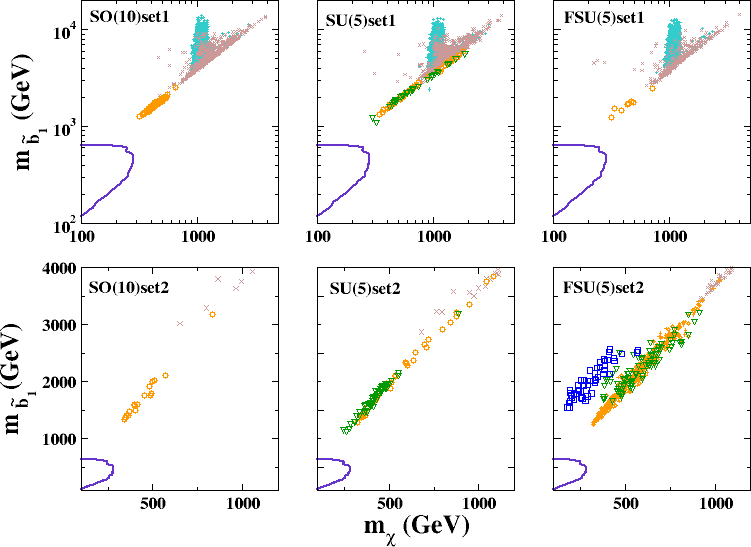}
\end{center}
\caption{ Scatter plots of non-universal GUT models in the ($m_\chi $, $m_{\tilde{b}_1}$) plane, with the same
legend as in Fig.~\ref{fig:GUTpar}. The solid purple line is the ATLAS 95 \%CL limit from~\cite{ATLAS_3rd_squarks_1,Aad:2013ija_3rd_squarks_2}. See the text for details.
}
\label{fig:msb1_mlsp}
\end{figure*}

As seen in Fig.~\ref{fig:mst1_mlsp}, the stop mass in the models we study is generally larger than 800 GeV,
and the present limits from searches in 
$\tilde{t}_1 \to t \chi^0_1$, do not reach such values. On the other hand, the projected sensitivity
with 3000 fb$^{-1}$ will partly cover the \stau\ coannihilation regions. 
The flipped SU(5) Set 2 scan displays the stop coannihilation strip where
both the $\tilde{t}_1$ and neutralino mass are in the range 200-600 GeV. However, we see that the \stop\ strip is not 
affected by the above-mentioned search, 
though we have already seen that it is constrained indirectly 
by the limits in Figs.~\ref{fig:mhf_m0}, \ref{fig:mg_mq} and \ref{fig:mch1_mlsp}.

We see in Fig.~\ref{fig:msb1_mlsp} that the lighter sbottom squark is heavier than
1 TeV in all the panels. The present 95 \%CL limit for $\tilde{b}_1$ pair production 
decaying to $b\chi$~\cite{ATLAS_3rd_squarks_1,Aad:2013ija_3rd_squarks_2} 
does not reach the favoured regions, and the searches for this sparticle are not competitive 
with the other channels.

\paragraph{Complementarity of searches:} Under the assumption that the lightest neutralino constitutes all the observed relic abundance, Figs.~\ref{fig:sigmaSI}, \ref{fig:sigmaann}, \ref{fig:mhf_m0} and \ref{fig:mg_mq}  show the complementarity 
of dark matter experiments and of LHC searches for 
supersymmetric particles. The GUT-inspired models and their respective
parameter spaces, as studied in our work, can be fully probed or excluded by combining  300 fb$^{-1}$ of data
accumulated by missing energy LHC searches (coannihilation areas), with the next generation of ton-scale direct-detection experiments. This is consistent with the results of \cite{Roszkowski:2014wqa},~\cite{Bagnaschi:2015eha}, where
similar complementarity was found in studies of the CMSSM, NUHM1, NUHM2 and pMSSM10.

\section{Summary and Conclusions}

In the following, we summarize the principal conclusions of this work.

\begin{itemize}

\item 
We have identified different patterns of soft SUSY-breaking terms at the GUT scale, depending
on the grand unification group, which we have used to distinguish different GUT scenarios via
their dark matter predictions and the constraints from LHC searches. 

\item 
We have calculated the SUSY spectra for the different gauge groups, finding
that the models predict different spectra for the same LSP mass,
connecting possible future observations with the structure of the
underlying unified theory.

\item
None of the GUT models studied offers high prospects for reducing substantially the
$a_\mu$ discrepancy via a SUSY contribution.

\item
In general, scenarios that favour degeneracies in the sparticle spectrum lead to
additional contributions to coannihilations, thus enhancing the efficiency and importance of these
processes.

\item We have studied the different relic density predictions 
and determined upper bounds for the neutralino mass in the different GUT scenarios.
We have also computed the cross sections for direct and indirect dark
matter detection in each case,  combining the bounds from
different dark matter experiments with those from LHC searches.

\item
We have found that SO(10), SU(5) and flipped SU(5) lead to very different predictions
for dark matter and LHC experiments, and
thus are distinguishable in future searches. Among other differences,
flipped SU(5)  predicts $\tilde{t}_1-\chi$ coannihilations that are absent  
in the other groups within the parameter ranges studied here, but can be explored by LHC searches.

\item
Direct searches for astrophysical dark matter scattering show interesting prospects for the Xenon 1 ton
and Darwin experiments, and models with a higgsino-like LSP or $A/H$ resonance annihilation may
offer prospects for future FERMI or CTA $\gamma$-ray searches. 

\item
The LHC searches for generic missing $E_T$, charginos and stops are quite complementary,
and future LHC runs will be able to constrain the models in several different ways.

\end{itemize}

The interesting prospects for exploring the parameter spaces of different SUSY GUT models found
in this paper, and the fact that their potential signatures are quite distinctive, whet our appetites
for data from LHC Run 2 and searches for astrophysical dark matter.

\section*{Acknowledgements}

The work of J.E. has been supported in part by the European Research Council via the Advanced 
Investigator Grant 267352, and by the UK STFC via the research grant ST/L000326/1.
The work of M.C. has been supported in part by the MINECO grant FPA2011-23778 and by
the MINECO/FEDER research grant CPI-14-397,
and acknowledges A. Pich, G. Rodrigo, J. Portol\'es,  P. Hern\'andez and N. Rius for hospitality at IFIC in Valencia 
where part of this work was done. M.C. and M.E.G. acknowledge support from the MINECO grant: "Fenomenolog\'{\i}a en F\'{\i}sica de 
Part\'{\i}culas y Astropart\'{\i}culas" (FPA2014-53631). S.L. thanks CERN for kind hospitality. 
R. RdA is supported by the Ram\'on y Cajal program of the Spanish MICINN, by
the Invisibles European ITN project (FP7-PEOPLE-2011-ITN, 
PITN-GA-2011-289442-INVISIBLES), and by the MEC projects
``SOM: Sabor y origen de la Materia" (2014-57816) and ``Fenomenolog\'ia y Cosmolog\'{\i}a de 
la F\'{\i}sica mas all\'a del Modelo Est\'{a}ndar e implicaciones Experimentales 
en la era del LHC" (FPA2013-44773).
M.C., M.E.G. and R.RdA acknowledge support from
Spanish MICINN Consolider-Ingenio 2010 Program under the grant MULTIDARK CSD2209-00064. 


\end{document}